\documentclass[referee]{raa}

\usepackage{graphicx,times}
\usepackage{natbib}
\usepackage{amssymb,amsmath}
\usepackage[T1]{fontenc}
\bibpunct{(}{)}{;}{a}{}{,}

\usepackage[pagebackref=true]{hyperref}

\begin{document}

   \title{Lightcurve Features of Magnetar-Powered Superluminous Supernovae with Gravitational-Wave Emission and \\ High-Energy Leakage}

\volnopage{ {\bf 20XX} Vol.\ {\bf X} No. {\bf XX}, 000--000}
	\setcounter{page}{1}

	\author{Jinghao Zhang
   		\inst{1,2}
   	\and Yacheng Kang
   		\inst{1,2}
   	\and Jiahang Zhong
   		\inst{1,2}
   	\and Hong-Bo Li
   		\inst{2}
   	\and Liang-Duan Liu
   		\inst{3,4}
   	\and Yun-Wei Yu
   		\inst{3,4}
   	\and Lijing Shao
   		\inst{2,5}
   }

	\institute{Department of Astronomy, School of Physics, Peking University, Beijing 100871, China; {\it yckang@stu.pku.edu.cn}\\
		\and
            Kavli Institute for Astronomy and Astrophysics, Peking University, Beijing 100871, China; {\it lihb2020@pku.edu.cn}; {\it lshao@pku.edu.cn}\\
		\and
			Institute of Astrophysics, Central China Normal University, Wuhan 430079, China\\
		\and
			Education Research and Application Center, National Astronomical Data Center, Wuhan 430079, China\\
		\and
			National Astronomical Observatories, Chinese Academy of Sciences, Beijing 100012, China\\
\vs \no
   {\small Received 20XX Month Day; accepted 20XX Month Day}
}

\abstract{Superluminous supernovae (SLSNe) are a distinct class of stellar explosions, exhibiting peak luminosities 10--100 times brighter than those of normal SNe. Their extreme luminosities cannot be explained by the radioactive decay of \(^{56}\mathrm{Ni}\) and its daughter \(^{56}\mathrm{Co}\) alone. Consequently, models invoking newly formed millisecond magnetars have been widely proposed, capable of supplying additional energy through magnetic dipole radiation. For these rapidly rotating magnetars, however, gravitational-wave (GW) emission may also contribute significantly to the spin-down, particularly during their early evolutionary stages. While high-energy photons initially remain trapped within the optically thick ejecta, they will eventually escape as the ejecta becomes transparent during the expansion, thereby influencing the late-time lightcurve. In this work, we adopt an analytical framework to systematically explore the combined effects of GW emission and high-energy leakage on the lightcurve of SLSNe. Compared to scenarios that neglect these processes, we find that for magnetars with initial spin periods of millisecond, the combined influence suppresses early-time luminosities but enhances late-time emission. We further investigate the effects of the neutron-star equation of state to the lightcurve, GW emission efficiency, ejecta mass, and other relevant quantities. Our results highlight the complex interplay between GW-driven spin-down and radiative transport in shaping the observable features of SLSNe, offering new insights into diagnosing the nature of their central engines.
\keywords{supernovae: general --- stars: magnetars --- gravitational waves --- radiative transfer}
}

   \authorrunning{J. Zhang, et al. }            
   \titlerunning{SLSN Lightcurve Features}  
   \maketitle

%
\section{Introduction}           
\label{sec:intro}

Core-collapse supernovae (SNe), the cataclysmic explosions marking the deaths of massive stars, are among the most energetic events in the Universe, briefly outshining their host galaxies. These events serve as natural laboratories for probing stellar evolution, core-collapse dynamics, and nucleosynthesis processes \citep{Woosley:1986ta, Woosley:1995ip, Arnett:1996ev, Woosley:2002zz, Janka:2006fh, Smartt:2009zr, 2017suex.book.....B}. Within the broad population of observed SNe, superluminous SNe (SLSNe) form a rare and exceptionally luminous subclass, typically defined by their peak absolute magnitude $M < -21$\,mag in the optical bands \citep{Gal-Yam:2012ukv, Gal-Yam:2018out, Inserra:2019ciq}.\footnote{With the increasing number of SN discoveries, it has become evident that many events fainter than $-21$\,mag still exhibit spectroscopic features that are characteristic of SLSNe, indicating that no strict magnitude boundary exists \citep{Moriya:2024gqt}. Consequently, events brighter than $\simeq -20$\,mag are sometimes also classified as SLSNe.} The implementation of unbiased SN surveys in the 2000s significantly increased the detection rate of SLSNe, enabling statistical studies and the development of detailed theoretical models \citep{2017hsn..book..431H, Moriya:2018sig, Chen:2021wuo, Nicholl:2021pou}. Notably, the extreme luminosities of SLSNe cannot be accounted for by the radioactive decay of \(^{56}\mathrm{Ni}\) and \(^{56}\mathrm{Co}\) alone \citep{Nicholl:2013esa}, implying the need for an additional energy source to power the lightcurves.

SLSNe are commonly divided into two spectroscopic subclasses: Type I (hydrogen-poor) and Type II (hydrogen-rich); see \citet{Inserra:2019ciq} and \citet{Moriya:2024gqt} for comprehensive reviews.
In at least some of hydrogen-rich Type II SLSNe (typically SLSNe IIn), the interaction between the rapidly expanding ejecta and a dense circumstellar medium (CSM) can plausibly account for the observed optical emission lines \citep{Chevalier2011,Ginzburg2012,Chatzopoulos2012,Chatzopoulos2013}. By contrast, in hydrogen-poor Type I SLSNe, the absence of strong CSM signatures in X-ray and radio observations challenges such CSM-interaction models \citep{Nicholl:2016ouk, Margutti:2017lyd}. Instead, current consensus favors a compact central engine that deposits energy internally. This interpretation is further supported by observations of excess ultraviolet (UV) continuum in a subset of Type I SLSNe \citep{Nicholl:2016fum}. Moreover, the nebular spectra of several SLSN events resemble those SNe associated with $\gamma$-ray burst (GRB), which are known to be engine-driven explosions \citep{Nicholl:2016btp, Jerkstrand:2016uyh, Nicholl:2018ies}.

Following the terminal collapse of a massive star, fallback accretion onto the newly formed black hole (BH) can drive relativistic jets or disk winds capable of injecting additional energy into the ejecta. However, the amount of accreted mass needed in this scenario is generally too large to account for Type I SLSNe \citep{Moriya:2018hkw, Moriya:2024gqt}. Instead, models involving newly formed millisecond magnetars have emerged as the most widely favored scenario \citep{Maeda:2007ck, Kasen:2009tg, Woosley:2009tu, Dessart:2012vc, Inserra:2013ila, Metzger:2015tra, Sukhbold:2016epl}. 
Millisecond magnetars are
strongly magnetized, rapidly rotating neutron stars (NSs), typically with surface magnetic field strength ${B \simeq 10^{14} \mathrm{~G}}$ and initial spin period on the millisecond timescale. They can inject substantial energy into the SN ejecta via magnetic dipole radiation \citep{Goldreich:1969sb}. This process can drive a relativistic electron-positron wind that cools through synchrotron and inverse Compton (IC) radiation, producing high-energy photons in the X-ray and $\gamma$-ray bands \citep{Goldreich:1969sb, Kotera:2013yaa, Metzger:2013kia, Yu:2019exl, Vurm:2021dgo}. These photons are initially trapped and thermalized by the optically thick ejecta, serving as an additional internal heat source that powers the SLSN emission.

Meanwhile, if a rapidly rotating magnetar is non-axisymmetric, its time-varying mass quadrupole moment would lead to the emission of gravitational waves \citep[GWs;][]{Zimmermann:1979ip, Jaranowski:1998qm, Nazari:2020ihd}. Such asymmetry may arise from static quadrupolar deformations, the so-called ``mountains'', in the outer crust of a magnetar \citep{Shapiro:1983du, Andersson:2019yve}, or from global inertial oscillation modes such as r-modes \citep{Chandrasekhar:1970pjp, Friedman:1978hf, Andersson:1997xt, Andersson:2000mf, Andersson:2019yve}. GW emission not only extracts a portion of the spin-down energy that would otherwise be deposited into the ejecta, but also accelerates the magnetar's rotational deceleration compared to a pure magnetic dipole braking, thereby significantly affecting the early-time evolution of the SLSN lightcurve \citep{Kashiyama:2015eua, Moriya:2016msa, Ho:2016qqm, Dai:2015jwa, Cheng:2018awt, Liu:2024fgt, Xie:2024gyo}. In addition, as the ejecta expands and becomes less dense, it gradually turns transparent to high-energy photons at late times, leading to some photon escape \citep{1997ApJ...491..375C, Chatzopoulos:2009uz, Chatzopoulos:2011vj, Kotera:2013yaa, Metzger:2013kia}. This reduces the thermalization efficiency and causes the observed SLSN lightcurve to deviate from prediction from the energy injection rate in the purely electromagnetically (EM) powered scenario \citep{2015MNRAS.452.1567C, Wang:2014cqa}. Observational signatures of this leakage effect have been reported in several SLSNe \citep{Nicholl:2016ouk, Lunnan:2016sqk, Nicholl:2017vde, Liu:2017upq, Lunnan:2017zna, Nicholl:2018cam, Zhu:2024yrc}.

In this work, we adopt an analytic framework to compute the bolometric lightcurves of magnetar-powered SLSNe, incorporating both GW emission and high-energy leakage effects. Compared with the purely EM-driven spin-down scenario, we find that for magnetars with  initial spin periods of millisecond, the combined influence of these two effects not only suppress the early-time luminosity, but also enhance the late-time emission---a behavior remains comparatively underexplored.\footnote{Note that \citet{Omand:2023fii} reported a similar behavior when varying the braking index of a magnetar, but they attributed it solely to differences in the energy-injection history, without further investigation. In contrast, we suggest that this feature is more naturally explained as a combined effect of the magnetar's energy injection and the thermalization efficiency of the ejecta.} The early-time suppression arises because part of the rotational energy is extracted by the GW emission rather than deposited into the ejecta, while the late-time enhancement results from a slower decline of the leakage effect owing to a reduced ejecta expansion rate when GW emission is present. Together, these effects also delay the lightcurve peak relative to the purely EM-driven spin-down case. We further find that this behavior is much less pronounced for magnetars with initial spin periods of $P_0 \gtrsim 10 \mathrm{~ms}$ or with weaker GW emission when the ellipticity $\varepsilon \lesssim 10^{-4}$ or the r-mode oscillation amplitude $\alpha \lesssim 0.01$. This indicates that only extremely fast-spinning magnetars with sufficiently strong GW emission can produce the diverse SLSN lightcurve evolution reported here. In addition, we examine the dependence of magnetar-powered SLSN lightcurves on the NS equation-of-state (EOS), as well as other parameters such as the magnetic field strength and ejecta mass. These results underscore the complex interplay between GW-driven spin-down and radiative transport in shaping the observable features of SLSNe, possibly offering new avenues for probing the nature of their central engines.

The structure of this paper is as follows. In Section~\ref{ sec:SLSNe_model }, we present the analytic formulation of SLSN lightcurves for different energy-loss scenarios, including treatments of GW emission and high-energy leakage. The resulting bolometric lightcurves and their comparisons are given in Section~\ref{ sec:result }. Finally, Section~\ref{ sec:conclusion } concludes the study.


\section{Magnetar-powered SLSNe}
\label{ sec:SLSNe_model }

In this section, we present in detail the framework of our analytic model for magnetar-powered SLSNe. Section~\ref{ subsec:magnetar } describes the spin-down evolution of a newly born magnetar in different energy-loss scenarios. Section~\ref{ subsec:leakage } introduces the treatment of high-energy leakage. Using these ingredients, Section~\ref{ subsec:lightcurve } outlines the computation of the bolometric SLSN lightcurve.


\subsection{Spin-down Evolution of a Millisecond Magnetar}
\label{ subsec:magnetar }

As mentioned in the Introduction, the spin-down of a rapidly rotating millisecond magnetar can be powered by two principal mechanisms: EM radiation, primarily through magnetic dipole emission, and GW emission, either from a quadrupolar deformation or from fluid oscillations \citep{Lasky:2015uia}. To facilitate a clear comparison of the relative contributions from different spin-down channels, we consider the following three cases:
\begin{itemize}
    \item \textbf{Case I}: Spin-down driven solely by EM radiation;
    \item \textbf{Case II}: EM radiation plus GW emission from a quadrupolar deformation characterized by an ellipticity $\varepsilon$;
    \item \textbf{Case III}: EM radiation plus GW emission from an r-mode oscillation with amplitude $\alpha$.
\end{itemize}

Specifically, the EM spin-down luminosity from magnetic dipole radiation can be written as \citep{Pacini:1968gfj, 1969Natur.221..454G, Spitkovsky:2006np, Contopoulos:2013ota},
\begin{equation}
    L_\mathrm{EM} = \frac{B^2 R^6 \sin^2\theta}{6 c^3} \Omega^4 \equiv \beta I\Omega^4 \,,
    \label{ eq:L_mag }
\end{equation}
where $c$ is the speed of light; $R$ is the magnetar redius; $B$ is the surface magnetic dipole field strength; $\theta$ is the angle between the rotation and magnetic axes; $\Omega = 2 \pi / P$ is the spin angular frequency for a spin period $P$. The parameter $\beta \equiv B^2 R^6 \sin^2\theta / 6 c^3 I$, where $I$ is the moment of inertia. For simplicity, we assume an orthogonal rotator throughout this paper, i.e., $\sin\theta = 1$.

For GW emission, the mountain-induced GW luminosity in Case II is \citep{Shapiro:1983du, Usov:1992zd, Zhang:2000wx, Ho:2016qqm},
\begin{equation}
    L_\mathrm{GW,e} = \frac{32 G I^2 \varepsilon^2}{5 c^5} \Omega^6 \equiv  \gamma_\mathrm{e} I \Omega^6 \,,
    \label{ eq:L_gw_e }
\end{equation}
where $G$ is the gravitational constant and ${\gamma_\mathrm{e} \equiv 32 G I \varepsilon^2 / 5 c^5}$. Alternatively, for r-mode-driven GW emission in Case III, its energy-loss rate can be expressed as \citep{1998PhRvD..58h4020O, Andersson:2000mf, Ho:2016qqm},
\begin{equation}
    L_\mathrm{GW,r} = \frac{131072 \pi G M R^4 \tilde{J}^2 I}{54675 c^7 \tilde{I}} \alpha^2 \Omega^8 \equiv  \gamma_\mathrm{r} I \Omega^8 \,,
    \label{ eq:L_gw_r }
\end{equation}
where $M$ is the magnetar mass, and 
\begin{equation}
    {\gamma_\mathrm{r} \equiv 
\frac{131072 \pi G M R^4 \tilde{J}^2}{54675 c^7 \tilde{I}} \, \alpha^2} \,.
\end{equation}
Following \citet{Ho:2016qqm}, we adopt $\tilde{J} \simeq 0.01635$ and $\tilde{I} \simeq 0.261$.

Given the EM and GW energy-loss rates for the three cases, the magnetar's spin-down evolution is then described via energy conservation,
\begin{equation}
\frac{\mathrm{d} E_{\mathrm{rot}}}{\mathrm{d} t} 
= I \Omega \frac{\mathrm{d} \Omega}{\mathrm{d} t}
= -L_{\mathrm{EM}}-L_{\mathrm{GW}} \,,
\label{ eq:dE_rot_dt }
\end{equation}
where $E_{\mathrm{rot}}=I \Omega^{2} / 2$ is the total rotational energy of the magnetar; $L_\mathrm{GW}$ denotes either $L_\mathrm{GW,e}$ (Case II) or $L_\mathrm{GW,r}$ (Case III). Substituting Equations~\eqref{ eq:L_mag }--\eqref{ eq:L_gw_r } into Equation~\eqref{ eq:dE_rot_dt } yields,
\begin{equation}
\begin{aligned}	
\frac{\mathrm{d} \Omega}{\mathrm{d} t} = 
\left\{ 
\begin{array}{lll}
-\beta \Omega^{3} \,, \hspace{1cm} 
& \text{for Case I}\,,\\
-\beta \Omega^{3}-\gamma_{\mathrm{e}} \Omega^{5} \,, \hspace{1cm} 
& \text{for Case II}\,,\\
-\beta \Omega^{3}-\gamma_{\mathrm{r}} \Omega^{7} \,, \hspace{1cm} 
& \text{for Case III}\,.
\end{array}
\right.
\end{aligned}
\label{ eq:dOmega_dt }
\end{equation}
Because GWs propagate unimpeded through the surrounding ejecta, the rotational energy they carry away cannot be used to power the SLSN emission. Thus, in Cases II and III, the coupled EM and GW losses in Equation~\eqref{ eq:dOmega_dt } jointly accelerate the magnetar spin-down and can significantly reduce the SLSN luminosity compared to the purely EM-driven spin-down scenario (Case I). 


\subsection{High-energy Leakage Effect}
\label{ subsec:leakage } 

For magnetar-powered SLSNe, most of the EM spin-down energy is expected to emerge as a Poynting-flux-dominated pulsar wind (PW), with X-ray and $\gamma$-ray photons produced through interactions between the PW and the surrounding ejecta \citep{Metzger:2013kia, Wang:2014cqa, Vurm:2021dgo}. During the first several months after the explosion---when the optical SLSN lightcurve is near its peak---IC scattering can dominate the production of high-energy photons, leading to a $\gamma$-ray-dominated spectrum \citep{Vurm:2021dgo}. Although the ejecta is initially dense enough to fully thermalize these photons via Compton downscattering, Coulomb collisions, and photoionization, the optical depth decreases as the ejecta expands. This reduction allows partial escape of hard photons, lowering the thermalization efficiency. Because X-rays are expected to remain trapped for a much longer timescale ($\simeq3$--$30$\,yr) than $\gamma$-rays \citep{Margalit:2018bje}, we follow previous studies in accounting only for $\gamma$-ray leakage in this work \citep{Wang:2014cqa, 2015MNRAS.452.1567C, Xie:2024gyo}. Worthy to mention that, although no such leakage has been directly detected in SLSNe, with only upper limits reported \citep{Margutti:2017lyd, Bhirombhakdi:2018hil, Renault-Tinacci:2017gon}, the late-time steepening observed in several SLSN lightcurves is more consistent with a contribution from $\gamma$-ray escape.

The leakage effect can be parameterized by a dimensionless trapping factor \citep{Wang:2014cqa},  
\begin{equation}
	\eta (t) \equiv 1 - \mathrm{e}^{-\tau_\mathrm{\gamma} (t)} \,,
	\label{ eq:eta }
\end{equation}
where $\eta (t)$ represents the fraction of high-energy photons thermalized within the ejecta. The $\gamma$-ray optical depth $\tau_\mathrm{\gamma} (t)$ in Equation~\eqref{ eq:eta } is given by,
\begin{equation}
\tau_\mathrm{\gamma}(t) = \frac{3 \kappa_\mathrm{\gamma} M_\mathrm{ej}}{4 \pi R_\mathrm{ej}^2} \simeq \frac{3 \kappa_\mathrm{\gamma} M_\mathrm{ej}}{4 \pi v^2} t^{-2} \,,
\label{ eq:tau_gamma }
\end{equation}
where $M_\mathrm{ej}$ and $R_\mathrm{ej}$ are the ejecta mass and radius, respectively; $\kappa_\mathrm{\gamma}$ is the $\gamma$-ray opacity. In this work, we adopt a fiducial gray opacity, $\kappa_\gamma \simeq 0.1 \mathrm{~cm^2 ~g^{-1}}$, the median value inferred from fits to Type I SLSNe by \citet{Nicholl:2017vde}.


\subsection{SLSN Lightcurve}
\label{ subsec:lightcurve }

To model the bolometric SLSN lightcurve, we adopt the one-zone approximation of \citet{Kasen:2009tg}, assuming spherical symmetry and uniform internal conditions within the ejecta. The internal energy $E_\mathrm{int}$ evolves according to the first law of thermodynamics \citep{Kasen:2015nma},
\begin{equation}
\frac{\mathrm{d} E_{\mathrm{int}}}{\mathrm{d} t} 
= \eta L_{\mathrm{EM}}-L_{\mathrm{bol}}-4 \pi R_\mathrm{ej}^{2} vp \,,
\label{ eq:dE_int_dt }
\end{equation}
where $\eta$ is the $\gamma$-ray trapping factor \eqref{ eq:eta }; ${v \equiv \mathrm{d} R_\mathrm{ej}/\mathrm{d} t}$ is the ejecta expansion velocity; $p$ is the radiation pressure, which is given as
\begin{equation}
	p = \frac{E_\mathrm{int}}{3 V} 
	= \frac{E_\mathrm{int}}{4 \pi R_\mathrm{ej}^3} \,.
	\label{ eq:p_rad }
\end{equation}
The ejecta dynamics follow Newton's second law,
\begin{equation}
	M_\mathrm{ej} \, \frac{\mathrm{d} v}{\mathrm{d} t} = 4 \pi R_\mathrm{ej}^2 p \,.
	\label{ eq:M_dv_dt }
\end{equation}
Note that $L_\mathrm{bol}$ in Equation~\eqref{ eq:dE_int_dt } is the radiated bolometric luminosity, which can be approximated by \citep{1980ApJ...237..541A, Arnett:1982ioj, Kasen:2009tg, Yu:2016ils}, 
\begin{equation}
L_{\mathrm{bol}} (t) \simeq \frac{c E_{\mathrm{int}}}{R_\mathrm{ej} \tau}\left(1-\mathrm{e}^{-\tau}\right) \,,
\label{ eq:L_bol }
\end{equation}
where $\tau =  3 \kappa_\mathrm{es} M_\mathrm{ej} / 4 \pi R_\mathrm{ej}^2$ is the optical depth of the ejecta. Throughout this work, we adopt a fiducial electron-scattering opacity of $\kappa_\mathrm{es} \simeq 0.1 \mathrm{~cm^2 ~g^{-1}}$. Using the above equations, we can then compute the time evolution of the observed bolometric SLSN lightcurve. We note that both $\kappa_\mathrm{es}$ and $\kappa_\gamma$ are assumed to be constant and gray in our simple model, whereas in reality they should vary with ionization state, temperature, and composition. More accurate modeling therefore requires detailed radiative transfer calculations, which are beyond the scope of this analytical framework and are deferred to a future study.

 
\section{Result}
\label{ sec:result }

\begin{figure}
\centering
\includegraphics[width=12cm]{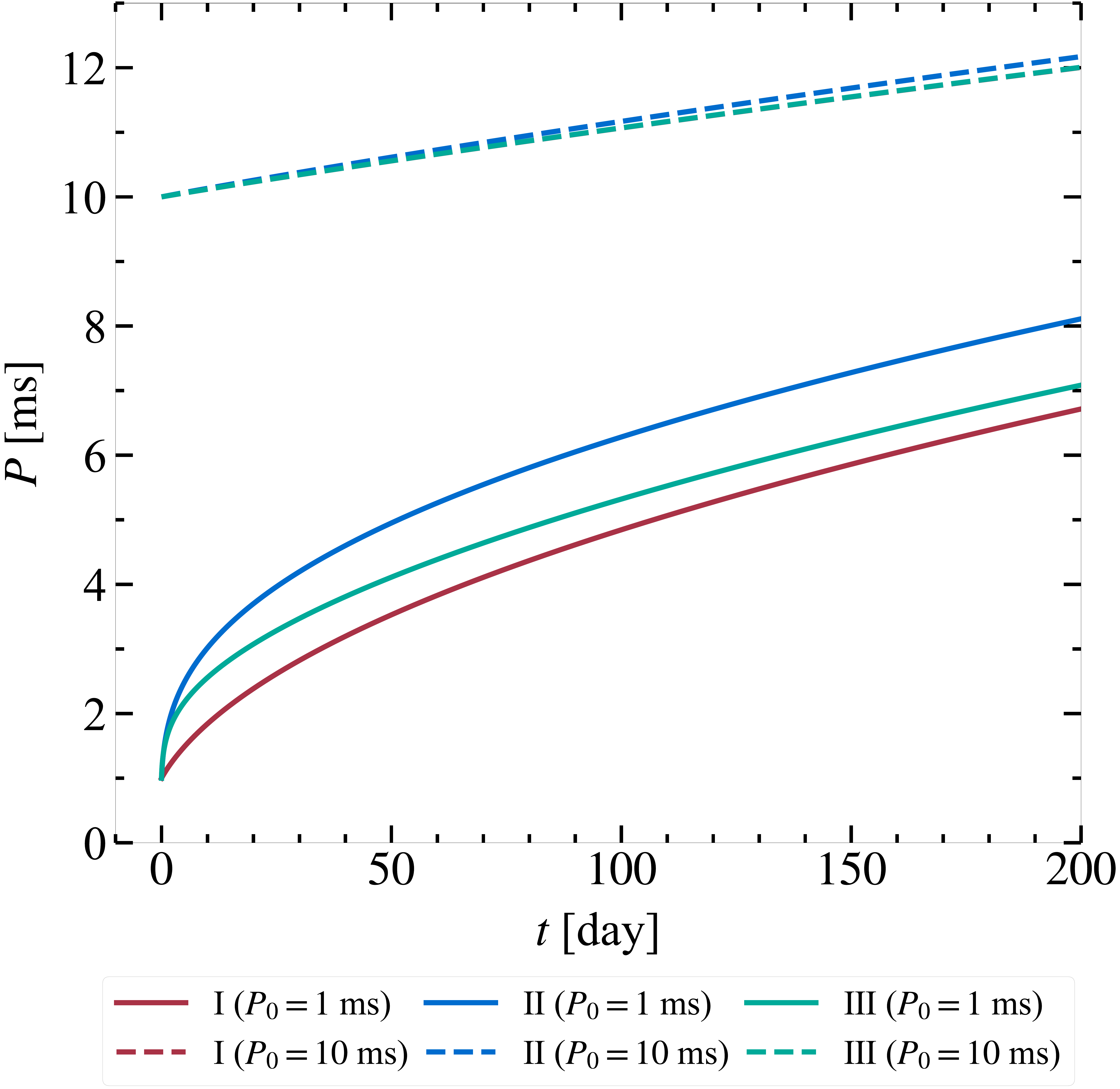}
\caption{Evolution of the magnetar spin period. Different colors denote different energy-loss scenarios (see Section~\ref{ subsec:magnetar }). We fix $\varepsilon = 10^{-3}$ for Case II and $\alpha = 0.1$ for Case III. Solid lines correspond to an initial spin period of $P_0 = 1 \rm ~ms$, while dashed lines correspond to $P_0 = 10\rm ~ms$. Other parameters are fixed at $M = 1.4 ~\rm M_\odot$, $R = 10 \mathrm{~km}$, $I = 10^{45} \mathrm{~g ~cm^2}$, and $B=1 \times 10^{14} \mathrm{~G}$. When $P_0 = 10\rm ~ms$, GW emission becomes negligible, causing the evolution of spin period nearly overlaps with each other.}
\label{ fig:period }
\end{figure}

\begin{figure}
\centering
\includegraphics[width=12cm]{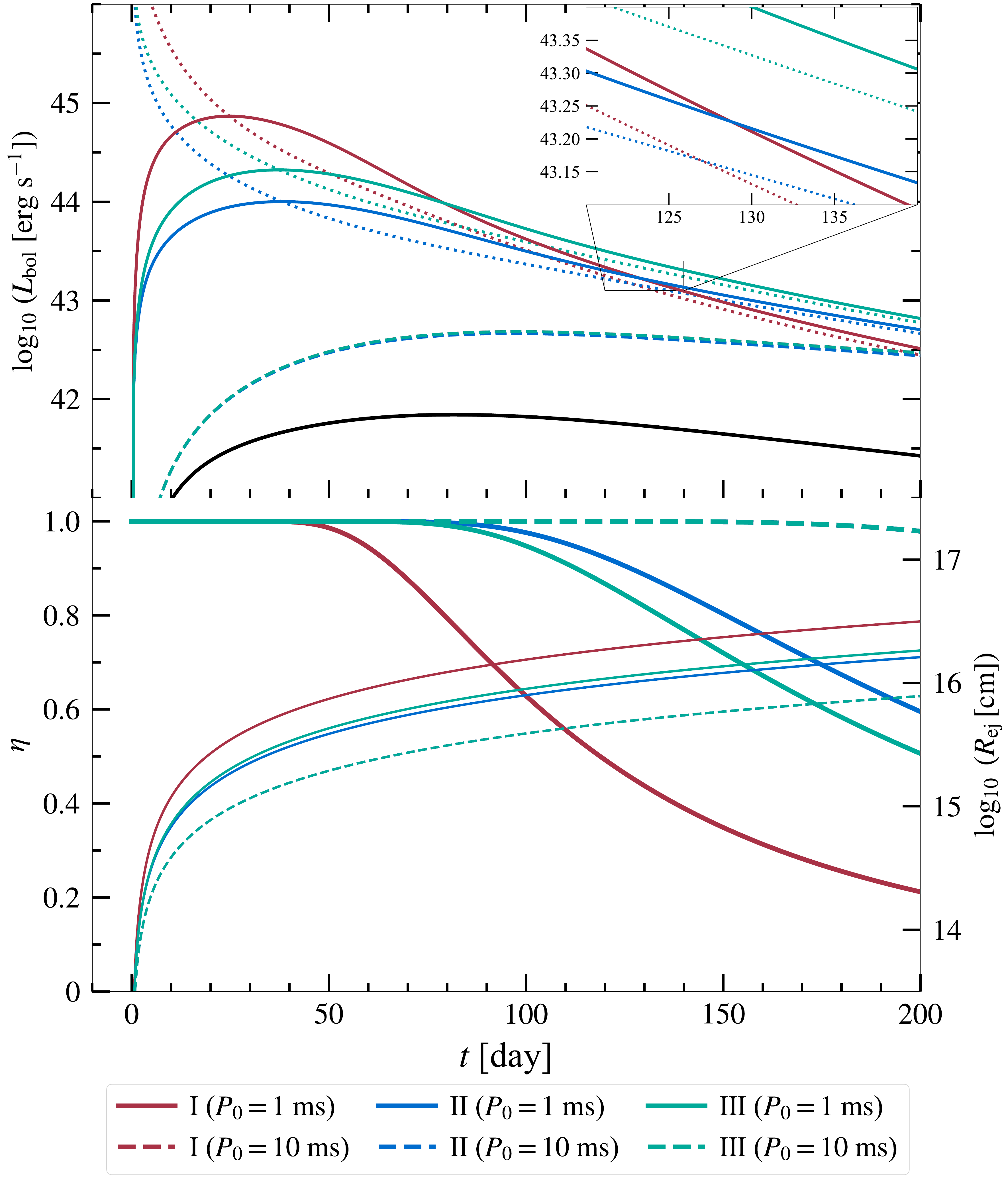}
\caption{The bolometric SLSN lightcurves (top) in different energy-loss scenarios, and the evolution of the $\gamma$-ray trapping factor $\eta$ and ejecta radius $R_\mathrm{ej}$ (bottom). Red, blue, and green curves correspond to Case I, II, and III, respectively. Solid lines represent an initial spin period of $P_0 = 1 \rm ~ms$, while dashed lines correspond to $P_0 = 10\rm ~ms$. In the top panel, dotted lines show the effective injected power, $\eta L_{\mathrm{EM}}$, for each case. The upper-right subpanel provides a zoomed-in view. For comparison, the black solid line shows the SN lightcurve powered solely by the radioactive decay of ${ }^{56} \mathrm{Ni}$ with a mass of $0.1\,\mathrm{M}_{\odot}$ \citep{1994ApJS...92..527N, Arnett:1996ev}. In the bottom panel, thick lines trace the evolution of $\eta(t)$ (vertical axis on the left), while thin lines indicate the ejecta expansion $R_{\mathrm{ej}}(t)$ (vertical axis on the right). Other parameters are fixed at $M = 1.4~\rm M_\odot$, $R = 10^6\mathrm{~cm}$, $I = 10^{45}\mathrm{~g\,cm^2}$, $B = 10^{14}\mathrm{~G}$, $M_{\mathrm{ej}} = 5~\rm M_\odot$, $\varepsilon = 10^{-3}$, and $\alpha = 0.1$. Some cases are indistinguishable in the plot for $P_0 = 10\,$ms. } 
\label{ fig:L_bol }
\end{figure}

We first present in Figure~\ref{ fig:period } the evolution of the magnetar spin period in different energy-loss scenarios. For an initial spin period of $P_0 = 1 \rm ~ms$, GW emission---present in either Case II or Case III---clearly dominates the early-time spin-down, increasing the spin-down rate when compared to the purely EM-driven spin-down case (Case I). At later times, the evolution is primarily governed by the magnetic dipole radiation. This behavior is consistent with the higher-order dependence of $\Omega$ in the GW spin-down term [see Equation~\eqref{ eq:dOmega_dt }], indicating that GW emission exerts a stronger influence during the early stages, when $\Omega$ is large. In contrast, for a slower initial spin period of $P_0 = 10 \rm ~ms$, Figure~\ref{ fig:period } shows negligible differences among the three cases.

Given the spin-down evolution in different energy-loss scenarios, the injected EM power can be computed from Equation~\eqref{ eq:L_mag }. Incorporating the $\gamma$-ray trapping factor $\eta$ and ejecta dynamics, we solve the coupled differential equations in Section~\ref{ subsec:lightcurve } to obtain the corresponding bolometric SLSN luminosities. Figure~\ref{ fig:L_bol } presents the bolometric SLSN lightcurves in the top panel for Cases I--III, and the evolution of $\eta$ and $R_\mathrm{ej}$ in the bottom panel. The top panel clearly shows that GW emission in either Case II or Case III substantially reduces the peak luminosity, consistent with the spin-down trends in Figure~\ref{ fig:L_bol }, as well as previous studies \citep{Ho:2016qqm, Cheng:2018awt, Liu:2024fgt, Xie:2024gyo}. This is expected, since GW radiation extracts part of the magnetar's rotational energy that would otherwise power the SLSN.

When $\gamma$-ray leakage effect is also included, however, a qualitatively different late-time trend emerges. Unlike the early-time divergence between energy-loss cases, Figure~\ref{ fig:L_bol } clearly shows that Case I exhibits the steepest post-peak decline and eventually its lightcurve becomes the faintest at late times---a behavior remains comparatively underexplored. This can be traced to the evolution of $\eta(t)$. For $P_0 = 1 \rm ~ms$, $\eta \simeq 1$ for all cases when $t \lesssim 50\,\mathrm{d}$, but beyond this point, $\eta$ in Case I drops first and remains the lowest thereafter, implying the lowest thermalization efficiency of high-energy photons and thus leading to the weakest late-time energy injection. This reduction in $\eta$ is directly linked to the more rapid ejecta expansion in Case I. As seen in the bottom panel of Figure~\ref{ fig:L_bol }, $R_\mathrm{ej}$ grows fastest in Case I due to the largest EM energy input, which significantly lowers the $\gamma$-ray optical depth $\tau_\mathrm{\gamma}$ [see Equation~\eqref{ eq:tau_gamma }] and hence $\eta$ [see Equation~\eqref{ eq:eta }]. Consequently, the combined effects of GW emission and high-energy leakage reshape the injected power as $\eta L_{\mathrm{EM}}$ [see Equation~\eqref{ eq:dE_int_dt }], suppressing early-time luminosity but enhancing late-time emission relative to the purely EM-driven spin-down scenario. These effects also delay the lightcurve peak. Notably, such behavior becomes much less pronounced for $P_0 = 10 \mathrm{~ms}$, indicating that only extremely rapidly spinning magnetars can produce the diverse lightcurve evolution discussed here.

\begin{figure}
\centering
\includegraphics[width=10cm]{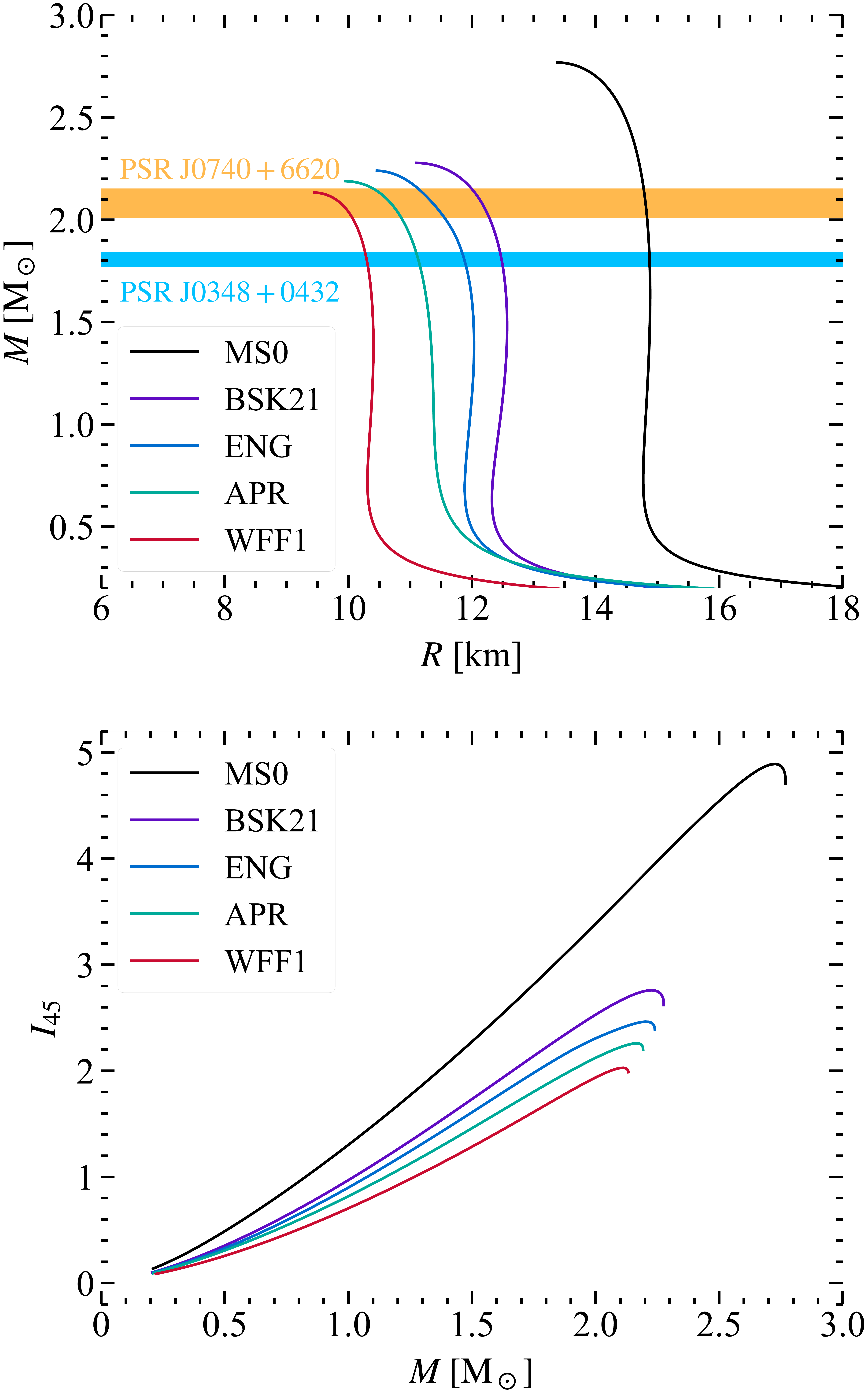}
\caption{The $M$--$R$ (top) and $I$--$M$ (bottom) relations for five representative NS EOSs: MS0, BSK21, ENG, APR, and WFF1 \citep{Lattimer:2000nx}. The 1-$\sigma$ confidence intervals of two precisely measured NS masses from PSRs~J0740+6620 \citep{Fonseca:2021wxt} and J0348+0432 \citep{Saffer:2024tlb} are shown as shaded regions in the top panel. In the bottom panel, the dimensionless moment of inertia $I_{45}$ is defined as $I_{45} \equiv I / 10^{45}~\mathrm{g\,cm^2}$.}
\label{ fig:M-RI }
\end{figure}

\begin{figure}
\centering
\includegraphics[width=12cm]{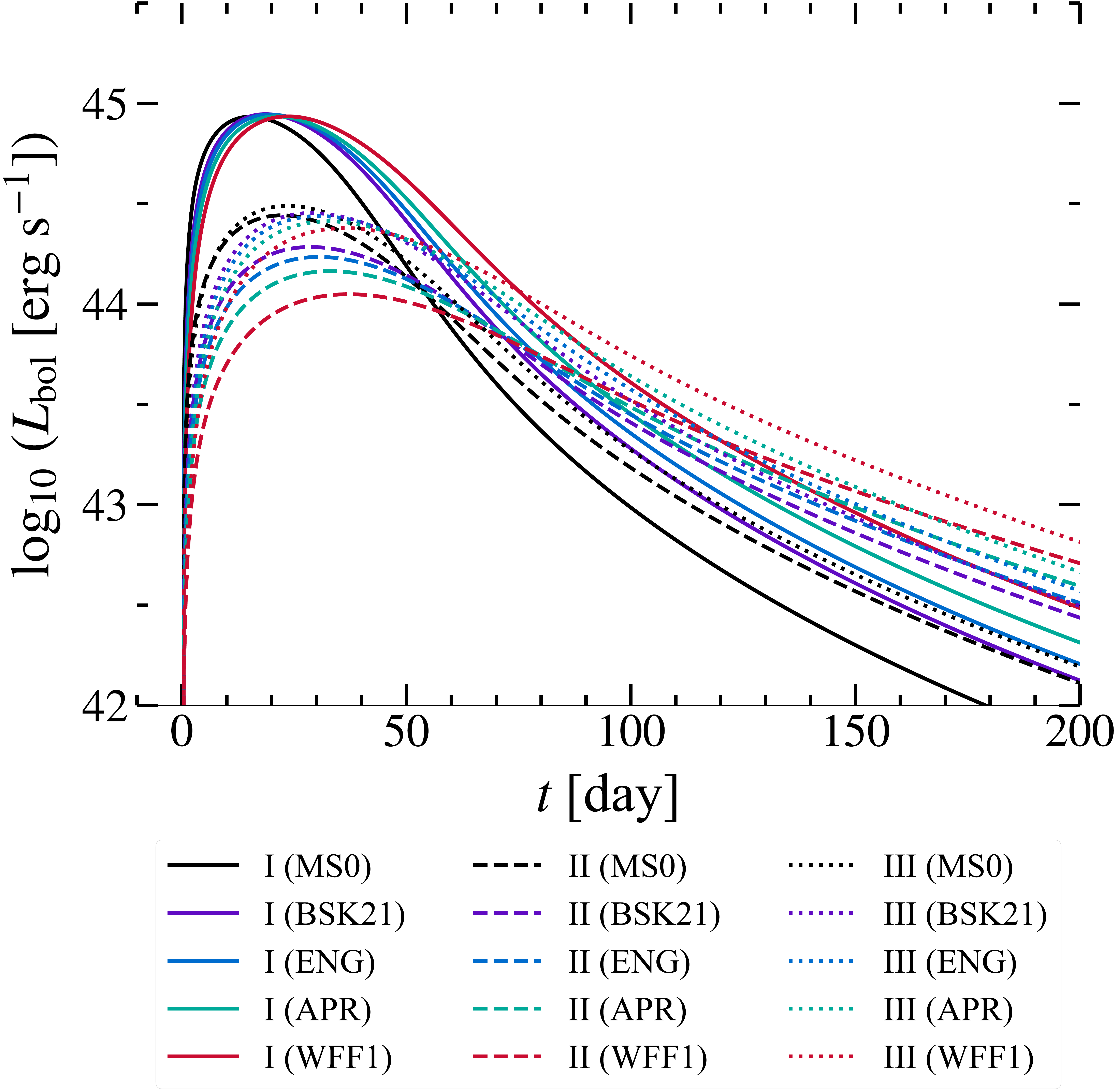}
\caption{Bolometric SLSN lightcurves for three spin-down scenarios with five different NS EOSs. Different colors represent different EOS models. Solid, dashed, and dotted lines correspond to Cases I, II, and III, respectively. Other parameters are fixed at $P_0 = 1\rm ~ms$, $M = 1.4~\rm M_\odot$, $M_{\mathrm{ej}} = 5~\rm M_\odot$, $B = 10^{14}\mathrm{~G}$, $\varepsilon = 10^{-3}$, and $\alpha = 0.1$.}
\label{ fig:EoS }
\end{figure}

\begin{figure}
\centering
\includegraphics[width=12cm]{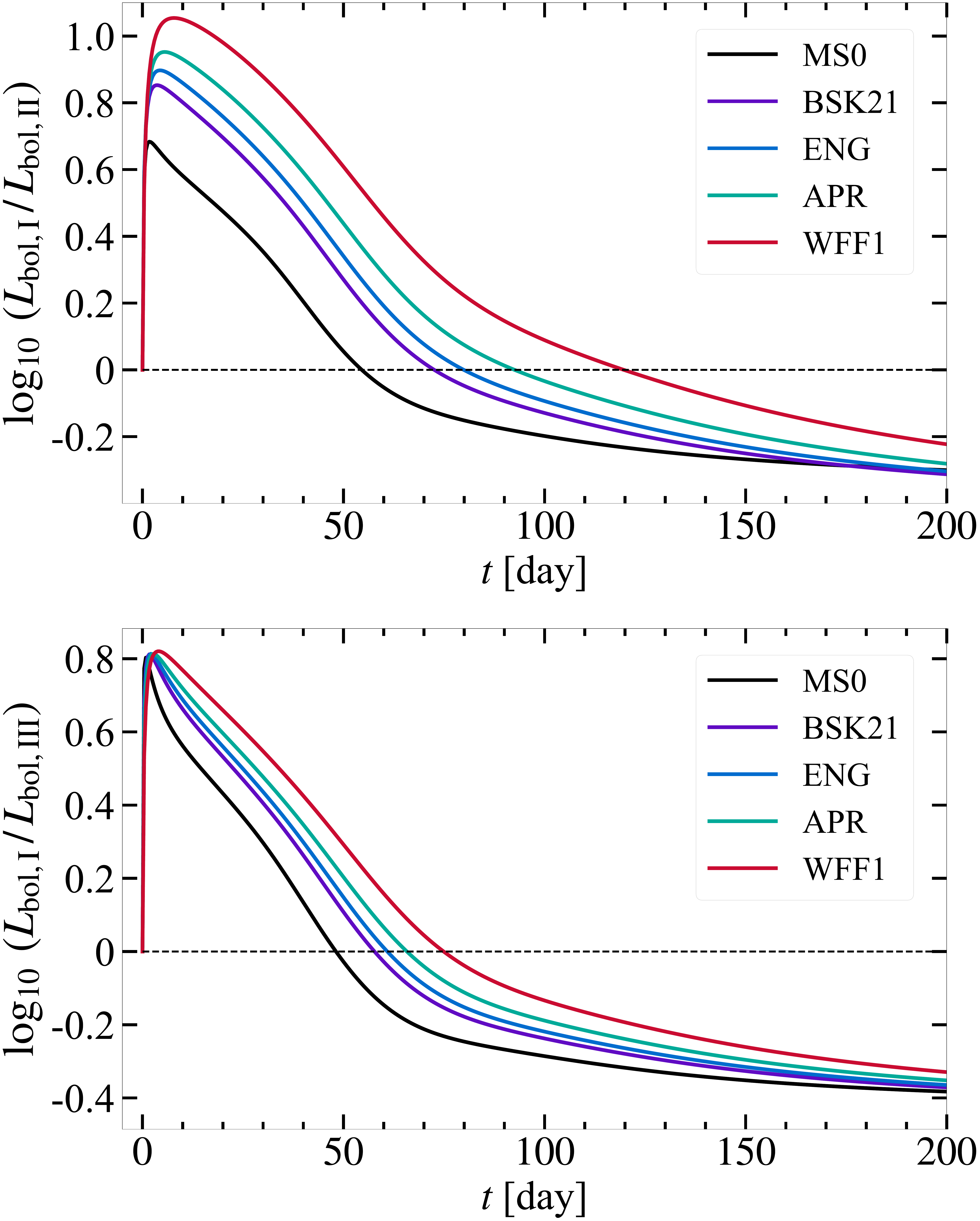}
\caption{Ratios of the bolometric SLSN luminosity between Case I and Case II (upper), and between Case I and Case III (lower). Different colors represent different EOS models. The horizontal dashed black line marks a  ratio of unity.}
\label{ fig:difference }
\end{figure}

The internal composition of NSs remains an open important question, and a wide range of candidate EOSs have been proposed, each predicting different mass--radius ($M$--$R$) and moment-of-inertia--mass ($I$--$M$) relations \citep[see e.g.,][]{Lattimer:2000nx, Gao:2021uus}. As shown in Section~\ref{ sec:SLSNe_model }, the spin-down evolution and the resulting bolometric lightcurve are sensitive to EOS-dependent magnetar properties, such as the radius $R$ and moment of inertia $I$. To assess the impact of the EOS, we consider five representative models: MS0, BSK21, ENG, APR, and WFF1 \citep{Lattimer:2000nx}. Figure~\ref{ fig:M-RI } shows their $M$--$R$ and $I$--$M$ relations. Incorporating this information, for each EOS we adopt precise values of $R$ and $I$ at a fixed magnetar mass of $M = 1.4 ~\rm M_\odot$ in the SLSN lightcurve calculations. The resulting bolometric SLSN luminosities for Cases I--III are presented in  Figure~\ref{ fig:EoS }, allowing a quantitative comparison. For the purely EM-driven spin-down case (Case I), Figure~\ref{ fig:EoS } indicates that different EOS models mainly affect the peak timescale, with little influence on the peak luminosity. In contrast, when GW emission is included in Case II or Case III, both the peak luminosity and the peak timescale exhibit significant EOS dependence. At late times, when the high-energy leakage effect becomes important, difference among NS EOS models can even alter the lightcurve evolution by up to an order of magnitude. This diversity underscores the need to account for magnetar structural properties when precisely modeling SLSN lightcurves.

\begin{figure}
\end{figure}

To further quantify these differences, Figure~\ref{ fig:difference } shows the ratios of the bolometric luminosity in Case I to that in Case II and Case III. The horizontal dashed black line marks a ratio of unity. At early times, the ratios are above unity, indicating that Case I produces higher luminosities than scenarios incorporating GW emission in Case II and Case III. However, due to the high-energy leakage, the luminosity in Case I eventually drops below that of Case II and Case III at late times. Notably, since $L_\mathrm{EM}$ exhibits a stronger dependence on $R$ ($\propto R^6$), an EOS with a larger $R$ enhances the role of EM radiation relative to GW emission, resulting in a lower ratio at early times. However, $L_\mathrm{GW}$ also increases with $R$ or $I$, so EOS with larger $R$ and $I$ ultimately produce stronger GW emission and yield a higher absolute ratio at late times. Furthermore, the transition epoch is EOS-dependent, with different models presenting distinct cross-over times. 

\begin{figure}[t]
\centering
\includegraphics[width=12cm]{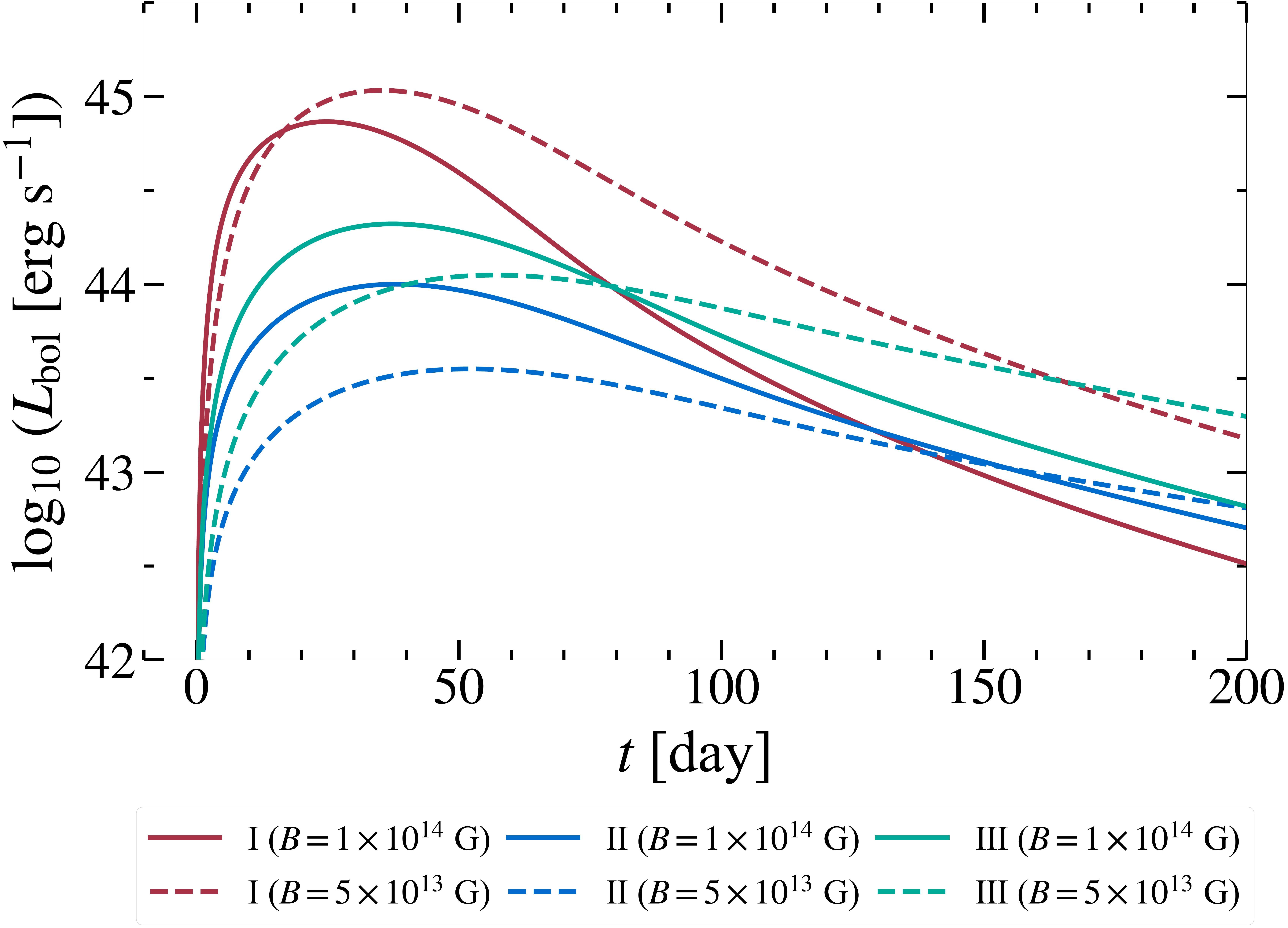}
\caption{Bolometric SLSN lightcurves for three spin-down cases with different magnetic field strengths. Red, blue, and green curves correspond to Case I, II, and III, respectively. Solid lines represent $B = 10^{14}\mathrm{~G}$, while dashed lines represent $B = 5 \times 10^{13}\mathrm{~G}$. Other parameters are fixed at $M_{\mathrm{ej}} = 5~\rm M_\odot$, $\varepsilon = 10^{-3}$, and $\alpha = 0.1$.}
\label{ fig:B_vary }
\end{figure}

\begin{figure}[t]
\centering
\includegraphics[width=11cm]{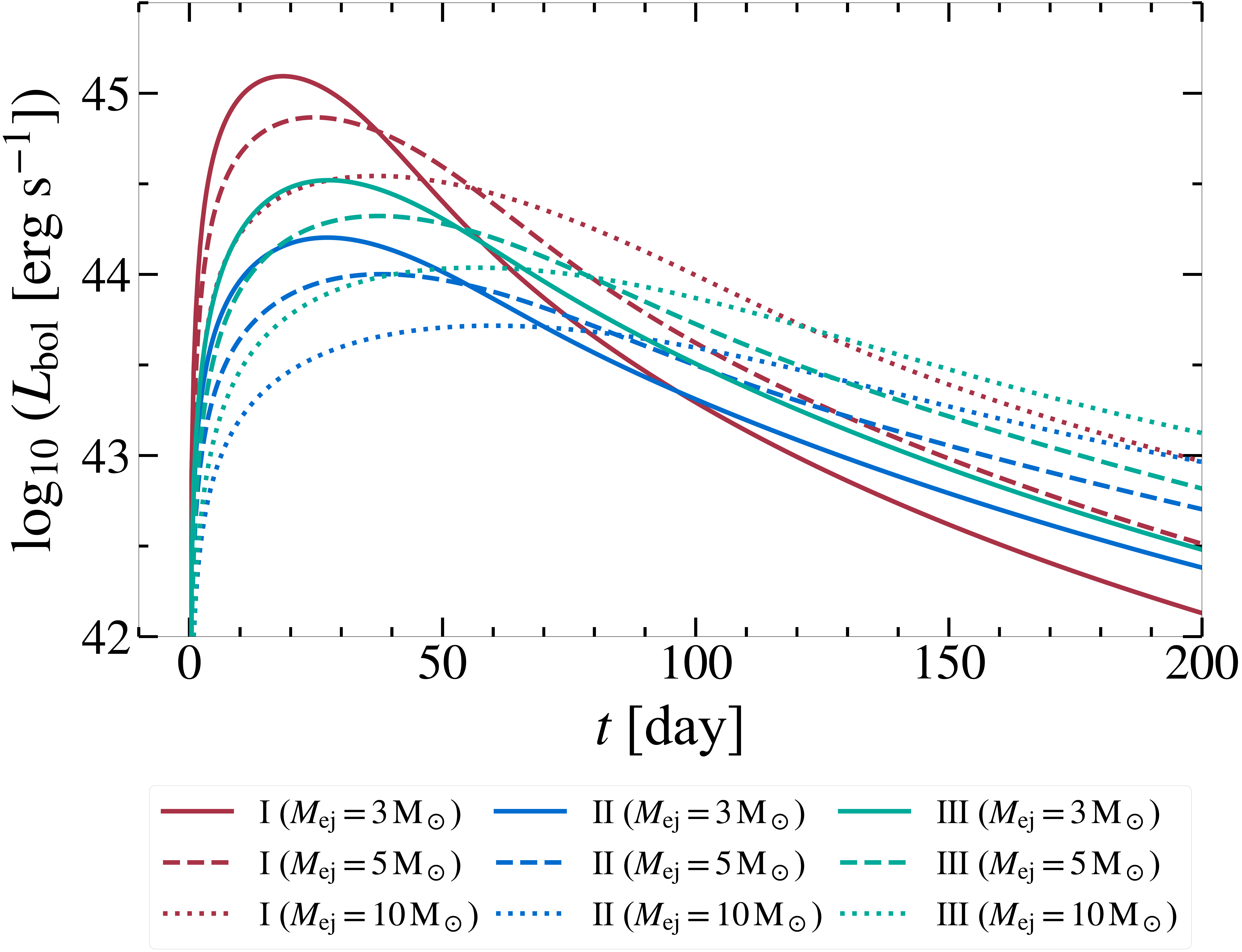}
\caption{Similar to Figure~\ref{ fig:B_vary }, but for different ejecta masses. Solid, dashed, and dotted lines represent $M_{\mathrm{ej}} = 3~\rm M_\odot$, $M_{\mathrm{ej}} = 5~\rm M_\odot$, and $M_{\mathrm{ej}} = 10~\rm M_\odot$, respectively. Other parameters are fixed at $B = 10^{14}\mathrm{~G}$, $\varepsilon = 10^{-3}$, and $\alpha = 0.1$.}
\label{ fig:M_ej_vary }
\end{figure}

\begin{figure}[t]
\centering
\includegraphics[width=11cm]{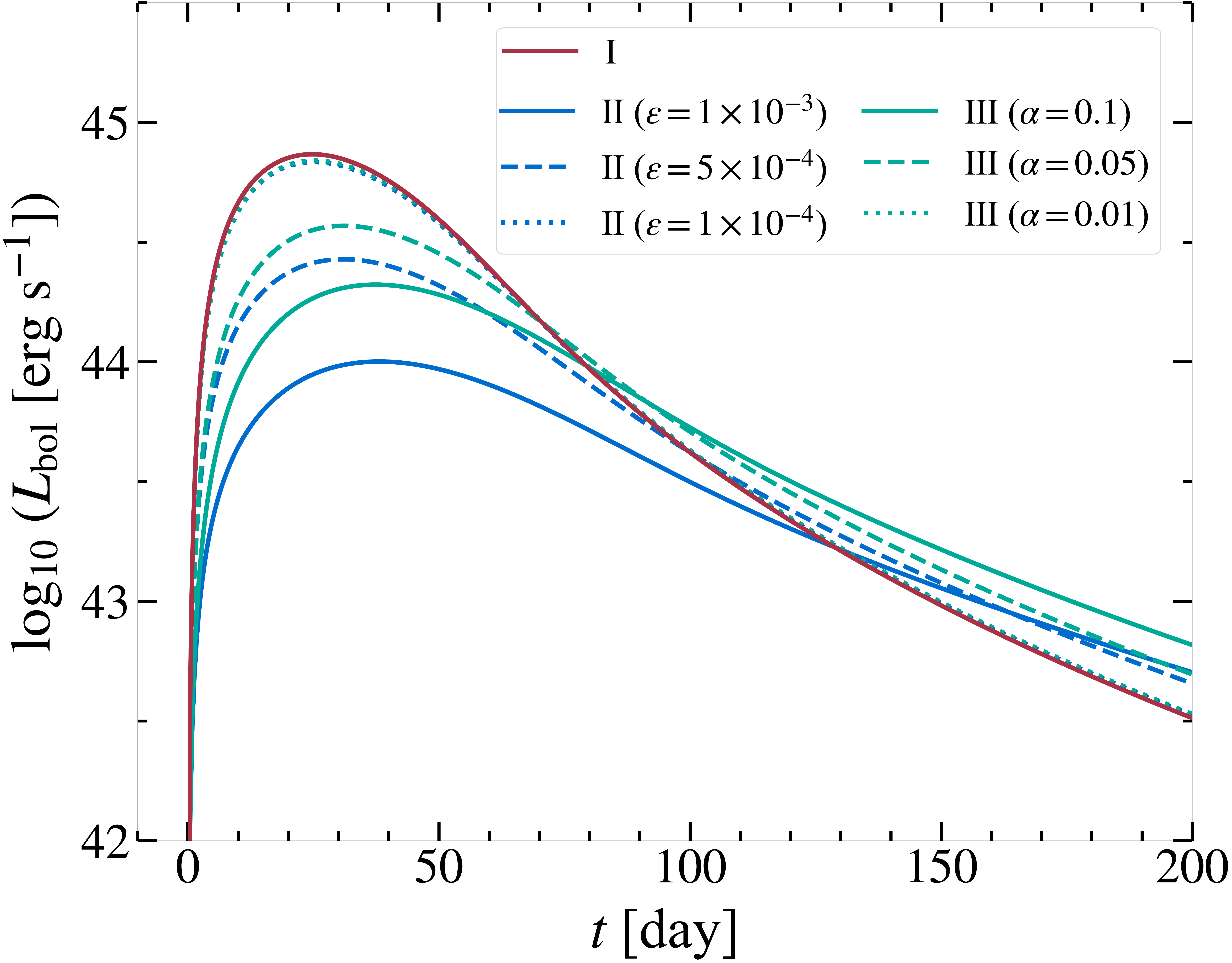}
\caption{Similar to Figure~\ref{ fig:B_vary }, but with line styles corresponding to different GW emission amplitudes for Cases II and III. Other parameters are fixed at $B = 10^{14}\mathrm{~G}$, and $M_{\mathrm{ej}} = 5~\rm M_\odot$.}
\label{ fig:GW_vary }
\end{figure}

In addition to the EOS dependence, we also explore the effects of varying other parameters, such as the magnetic field strength $B$, ejecta mass $M_{\mathrm{ej}}$, and GW emission amplitudes $\varepsilon$ and $\alpha$. These results are reported in Figure~\ref{ fig:B_vary } for $B$, in Figure~\ref{ fig:M_ej_vary } for $M_{\mathrm{ej}}$, and in Figure~\ref{ fig:GW_vary } for $\varepsilon$ and $\alpha$. In all these figures, the following parameters are fixed: $P_0 = 1\rm ~ms$, $M = 1.4~\rm M_\odot$, $R = 10^6\mathrm{~cm}$, and $I = 10^{45}\mathrm{~g\,cm^2}$.
Notably, for the purely EM-driven spin-down scenario (Case I) in Figure~\ref{ fig:B_vary }, a lower magnetic field of $B = 5 \times 10^{13}\mathrm{~G}$ yields a higher peak luminosity than $B = 1 \times 10^{14}\mathrm{~G}$, consistent with the scaling relation from \citet{Kasen:2009tg},
\begin{equation}
L_\mathrm{peak} \sim \frac{E_\mathrm{rot}t_\mathrm{EM}}{t_\mathrm{d}^2} \propto B^{-2} M_\mathrm{ej}^{-3/2} \kappa_{\mathrm{es}}^{-1} \,,
\label{ eq:L_peak }
\end{equation}
where $t_{\mathrm{EM}} \simeq P_0^2 / 4 \pi^2 \beta$ is the spin-down timescale and $t_{\mathrm{d}} \simeq \left(\kappa_{\mathrm{es}} M_{\mathrm{ej}} / v c\right)^{1 / 2}$ is the photon diffusion timescale. This relation applies when ${t_{\mathrm{EM}} \lesssim t_\mathrm{d}}$, requiring 
\begin{equation}
    B \gtrsim 2 \times 10^{13} \, \mathrm{G}
     \left( \frac{P_{0}}{1\,\mathrm{ms}} \right)
     \left( \frac{5 \, \rm M_\odot}{M_{\mathrm{ej}} } \right)^{3/8}
      \left( \frac{ 0.1 \, \mathrm{cm^2 \,g^{-1}} }{ \kappa_{\mathrm{es}} } \right)^{1 / 4} \,. \nonumber
\end{equation}
When GW emission is included in Cases II and III, this scaling no longer holds, as part of the rotational energy is lost to GWs rather than being deposited into the ejecta [see Equation~\eqref{ eq:dE_rot_dt }].
Figure~\ref{ fig:M_ej_vary } shows that increasing $M_{\mathrm{ej}}$ reduces the peak luminosity and delays the peak epoch, consistent with the longer diffusion time and the scaling relation \eqref{ eq:L_peak }. Figure~\ref{ fig:GW_vary } illustrates the effect of varying $\varepsilon$ and $\alpha$, showing that when $\varepsilon \lesssim 10^{-4}$ or $\alpha \lesssim 0.01$, GW losses become negligible, and the SLSN lightcurve evolution in Cases II and III effectively converges to that of the purely EM-driven spin-down scenario (Case I).
Overall, our findings emphasize the complex interplay between GW-driven spin-down and radiative transport in shaping the observable features of SLSN lightcurves, offering potential clues to the nature of their central engines. 


\section{Conclusion and Discussion}
\label{ sec:conclusion }

In this work, we adopt an analytic framework to compute the bolometric lightcurves of magnetar-powered SLSNe, incorporating both GW emission and high-energy leakage effects. Compared to the purely EM-driven spin-down scenario, we find that for magnetars with millisecond initial spin periods, the combined influence of these effects suppresses the early-time luminosity but enhances the late-time emission. The reduction in early-time luminosity arises because that a fraction of the rotational energy is lost to GW emission rather than being deposited into the ejecta. In contrast, the late-time luminosity enhancement---a behavior remains comparatively underexplored---is caused by a slower decline in the $\gamma$-ray trapping factor; this is a consequence of the reduced ejecta expansion rate when GW emission is present. Together, these effects also delay the lightcurve peak relative to the purely EM-driven spin-down case. Notably, such behavior becomes much less pronounced for magnetars with initial spin periods of $P_0 \gtrsim 10 \mathrm{~ms}$ or with weaker GW emission, say, ellipticity $\varepsilon \lesssim 10^{-4}$ or r-mode oscillation amplitude $\alpha \lesssim 0.01$. This indicates that only extremely fast-spinning magnetars with sufficiently strong GW emission can produce the diverse SLSN lightcurve evolution reported here. We further explore the dependence of magnetar-powered SLSN lightcurves on the NS EOS, as well as on other relevant parameters such as the magnetic field strength and ejecta mass. Our results highlight the complex interplay between GW-driven spin-down and radiative transport in shaping the observable features of SLSNe, possibly offering new avenues for diagnosing the nature of their central engines.

It is also worth emphasizing that the properties of a newly born magnetar following a core-collapse SN, and its surrounding environment, may differ substantially from those of older, isolated pulsars. In light of this, although we adopt $\varepsilon \lesssim 1 \times 10^{-3}$ and $\alpha \lesssim 0.1$ in this work---values consistent with previous studies \citep{Moriya:2016msa, Ho:2016qqm}---the actual upper limits of these parameters remain uncertain and merit further investigation. For instance, high-energy observations of X-ray plateaus following GRB events can place meaningful constraints on the ellipticity of such newborn magnetars \citep{Lasky:2015olc, Margalit:2017oxz, Xie:2022igk}. Although the current sensitivity of EM and GW observatories makes direct detection of associated signals from these magnetar-powered SLSNe unlikely, more sensitive facilities might provide crucial observational constraints in the near future \citep{Kashiyama:2015eua, Margalit:2017oxz, Villar:2018toe, Xie:2024gyo}. For example, the upcoming Legacy Survey of Space and Time (LSST), to be conducted by the Vera C. Rubin Observatory, is expected to discover $\mathcal{O}(10^4)$ SLSNe annually \citep{Villar:2018toe, LSST:2008ijt}. Such large datasets will offer unprecedented opportunities to probe the physics of magnetar-powered explosions, the diversity of SLSN lightcurve features, and the properties of their central engines.


\normalem
\begin{acknowledgements}

We are grateful to Yong Gao for sharing his code to calculate neutron star structure, and to Zhuo Li, Jin-Ping Zhu, Xishui Tian, Qinyuan Zhang, Qiang Wang, and Ruize Shi for insightful discussions. This work was supported by the National SKA Program of China (2020SKA0120300), the Beijing Natural Science Foundation (QY25099, 1242018), the National Natural Science Foundation of China (12447148, 12573042), the China Postdoctoral Science Foundation (2024M760081), the Max Planck Partner Group Program funded by the Max Planck Society, and the High-Performance Computing Platform of Peking University. YK is supported by the China Scholarship Council (CSC). LL is supported by the National Natural Science Foundation of China (grant Nos 12303047), Natural Science Foundation of Hubei Province (2023AFB321).

\end{acknowledgements}


\bibliographystyle{raa}
\bibliography{SLSN.bib}

\end{document}